\documentclass[conference]{IEEEtran}

\usepackage{graphicx}
\usepackage{booktabs}
\usepackage{amsmath}
\usepackage{amssymb}
\usepackage{multirow}
\usepackage{xcolor}
\usepackage{dblfloatfix}
\usepackage{balance}
\usepackage[caption=false,font=footnotesize]{subfig}
\usepackage{url}
\usepackage[hidelinks]{hyperref}

\title{Interpreting SSL Representations for Spoof Detection: a WavLM Study}

\author{
{
\centering
\fontsize{10}{10}\selectfont
\begin{tabular}{ccc}
\begin{tabular}[t]{c}
\textbf{Mohamed Mallat}\\
\textit{EURECOM}\\
Sophia Antipolis, France\\
Mohamed.Mallat@eurecom.fr
\end{tabular}
&
\begin{tabular}[t]{c}
\textbf{Michele Panariello}\\
\textit{EURECOM}\\
Sophia Antipolis, France\\
Michele.Panariello@eurecom.fr
\end{tabular}
&
\begin{tabular}[t]{c}
\textbf{Massimiliano Todisco}\\
\textit{EURECOM}\\
Sophia Antipolis, France\\
Massimiliano.Todisco@eurecom.fr
\end{tabular}
\vspace{0.25cm}
\\[1.2em]
\multicolumn{3}{c}{
\begin{tabular}{cc}
\begin{tabular}[t]{c}
\textbf{Nicholas Evans}\\
\textit{EURECOM}\\
Sophia Antipolis, France\\
Nicholas.Evans@eurecom.fr
\end{tabular}
&
\hspace{1.5cm}
\begin{tabular}[t]{c}
\textbf{Anthony Larcher}\\
\textit{Université du Mans}\\
Le Mans, France\\
Anthony.Larcher@univ-lemans.fr
\end{tabular}
\end{tabular}
}
\end{tabular}
}}

\begin{document}
\maketitle

\begingroup
\makeatletter
\renewcommand\@makefnmark{}
\renewcommand\@makefntext[1]{\noindent #1}
\makeatother

\footnotetext[0]{%
  \noindent\rule{\columnwidth}{0.4pt}\vspace{0.5ex}
  
  This work was supported by the French Agence Nationale de la Recherche (ANR) via the COMPROMIS (ANR-22-PECY-0011) project.
}
\endgroup

\begin{abstract}
Self-supervised speech models such as WavLM achieve strong spoofing detection performance, yet which layers encode the most spoof-relevant information and what acoustic properties they capture remain opaque.
We present a layer-wise interpretability study of WavLM using the ASVspoof~5 database, combining three complementary analyses: 
(i)~per-layer linear probing with attack-level evaluation across speech synthesis, voice conversion, and adversarial attacks; 
(ii)~a canonical correlation analysis (CCA) linking layer representations to spoofing-relevant acoustic features; (iii)~an exhaustive grid search over layer subsets showing that use of five layers outperforms full thirteen-layer pooling.
CCA analysis reveals that, while most acoustic features peak in alignment with mid-layers and then decay, voice quality measures jitter, shimmer, and the harmonics-to-noise ratio show the opposite trend, with correlation increasing toward later layers. 
Embedding visualisations reveal that deeper layers capture a continuous acoustic similarity among attacks rather than discrete categories, suggesting these layers encode fundamental synthesis artifacts rather than attack-specific signatures.
\end{abstract}

\section{Introduction}
Self-supervised learning (SSL) front-ends have become the backbone of modern spoofing countermeasures~(CMs) and presentation attack detection (PAD) solutions, achieving state-of-the-art performance for ASVspoof benchmarks~\cite{wang2024asvspoof5,jung2024espnet}. 
However, these models remain opaque; it is unclear which layers encode spoof-relevant information, how layer behaviour varies across attack types, and whether the learned representations can be linked to interpretable acoustic properties.  
Herein lie the goals of our work reported in this paper.
Recent work shows that spoofing countermeasures can exploit spurious cues and learning shortcuts~\cite{muller2021silence}, e.g.\ those linked to non-speech duration. This further motivates analyses that ground model behaviour in meaningful acoustic properties.

Among a plethora of available and popular alternative SSL models, we focus on WavLM~\cite{chen2022wavlm} which incorporates a denoising pre-training objective that exposes the model to noisy and overlapped speech. 
Because WavLM is trained to recover masked targets for a designated main speaker in noisy/overlapped mixtures, its denoising objective encourages representations that preserve speaker-related information and remain stable under acoustic distortions.

We report an investigation performed using the most recent ASVspoof~5 database~\cite{wang2025asvspoof5db}.
It provides a diverse set of bona fide speech utterances together with a set of spoofed/deepfake attacks generated using text-to-speech~(TTS), voice conversion~(VC), and adversarial perturbation methods, all treated with varied codecs and recording conditions~\cite{wang2024asvspoof5}. 
Its crowdsourced nature and the diversity of the data and acoustic conditions make it an ideal test bed for interpretability-driven analysis. 
We address three research questions:
\begin{enumerate}
    \item Which WavLM layers carry the most discriminative bona fide-spoof information?
    \item Does the detection of different attack families (TTS, VC, adversarial) depend upon the capture of information at different layer depths?
    \item What interpretable acoustic properties do WavLM layers encode, and how do these relate to known spoofing markers?
\end{enumerate}

\noindent The contributions of the work reported in this paper include:
\textbf{Attack-aware layer characterisation -} We provide a comprehensive analysis of how spoofing discriminability varies with representation depth and attack types, revealing strong layer-attack interactions and showing that mid-to-deep layers carry the most reliable bonafide/spoof cues under most conditions. \\
\textbf{Layer subset search -} We show that an optimally chosen compact layer subset can outperform standard all-layer pooling with a linear backend, providing a practical efficiency gain.\\
\textbf{CCA-based acoustic grounding -} We link layer representations to interpretable spoofing-relevant acoustic cues by computing projection-weighted canonical correlation analysis (PWCCA) showing a shift in representation geometry helping explain why certain layers behave differently and when they fail. t-SNE visualisation corroborates this: representations transition from discrete attack clusters (mid-layers) to a continuous horseshoe manifold (deep layers).

\begin{figure*}[!t]
\centering
\includegraphics[width=0.54\textwidth]{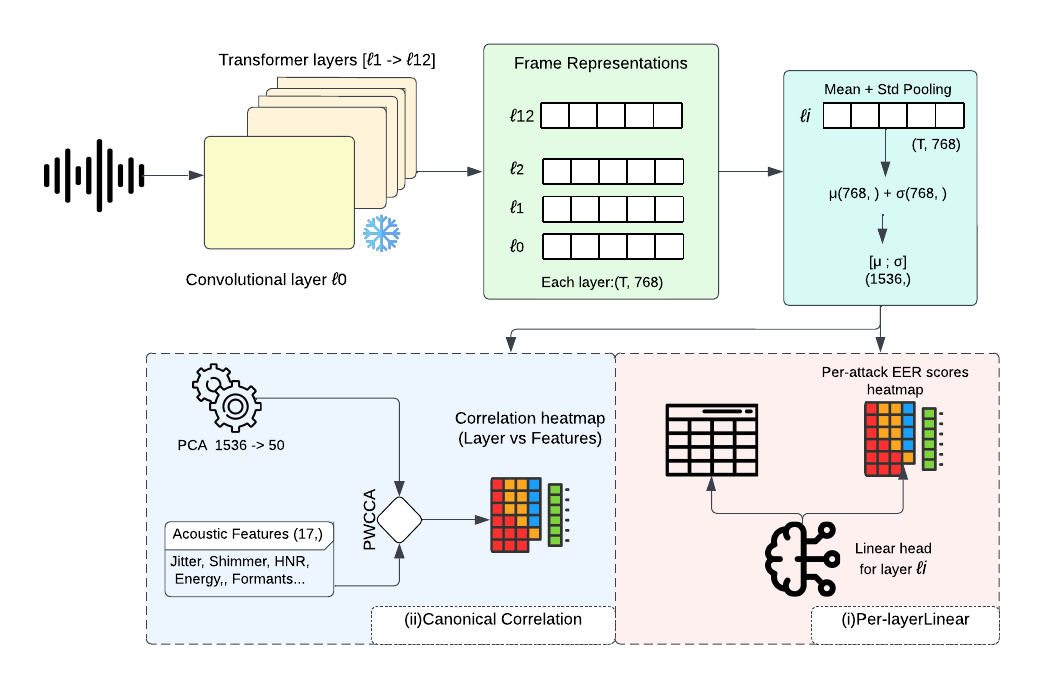}
\vspace{-0.55cm}\caption{\textbf{Overview of the analysis pipeline}. Top: WavLM produces frame‑level layer representations; each layer is summarised by mean+std pooling to a 1536‑D utterance vector. Bottom left (blue): PWCCA pipeline, PCA (1536→50) and correlation between pooled layer vectors and handcrafted acoustic features, producing layer vs feature heatmaps. Bottom right (pink): per‑layer linear probing, yielding per‑layer EER and per‑attack EER heatmaps.}
\label{fig:methodology}
\end{figure*}

\section{Related Work}

\subsection{Spoofing CMs and SSL front-ends}
Neural spoofing countermeasures (CMs) now span a broad range of back-ends, including graph-attention architectures such as AASIST, conformer-based anti-spoofing classifiers, and fused/ensemble systems~\cite{jung2022aasist,rosello2023conformer,wang2024asvspoof5,stourbe2024wavlm,zhu2024slim}. In recent ASVspoof 5 submissions, these back-ends are frequently paired with SSL front-ends such as wav2vec 2.0 and WavLM, often together with augmentation (e.g., RawBoost) and score/model fusion strategies~\cite{wang2024asvspoof5,wang2022sslcm,baevski2020wav2vec,chen2022wavlm,tak2022rawboost,stourbe2024wavlm}.
\vspace{-0.3cm}
\subsection{Layer-wise analysis of SSL models}                                Previous layer-wise studies of SSL speech representations~\cite{pasad2021layerwise,pasad2023comparative} show that early layers capture local acoustic features while late layers encode more abstract content and speaker information. In the context of spoofing detection, studies have explored SSL front ends~\cite{stourbe2024wavlm,wang2022sslcm,zhu2024slim}, but primarily to optimize detection performance. More recently, Xiao et al.~\cite{xiao2025layerwise} proposed layer-wise decision fusion for fake audio detection, demonstrating that per-layer decisions outperform feature-level fusion. 
Representation analysis tools such as singular vector canonical correlation analysis (SVCCA)~\cite{raghu2017svcca} and projection-weighted canonical correlation analysis (PWCCA)~\cite{morcos2018pwcca} provide quantitative similarity measures between representations. However, these tools have not yet been applied to interpret what spoofing-relevant information individual layers encode in an attack-dependent manner.
\vspace{-0.1cm}
\subsection{Acoustic cues for spoofing detection}
Specific acoustic features like jitter, shimmer, and the harmonic-to-noise ratio measures of vocal fold micro-perturbations are known markers of spoofed speech; TTS vocoders produce unnaturally smooth or irregular voice quality patterns~\cite{gupta2025pitchimperfect}. 
Formant trajectories have been shown to expose voice conversion artifacts; generative models struggle to replicate fine-grained articulatory detail~\cite{cuccovillo2024formant,borrelli2025forensic}. 
Spectral features such as sub-band centroids and flux have also been studied; spectral discontinuities from frame-level synthesis leave detectable traces in these measures~\cite{sahidullah2015comparison}. 
While none of these features alone match the detection performance of modern SSL-based systems, they offer the advantage of direct interpretability. The shift toward self-supervised models has improved accuracy but moved the field further from understanding what information drives detection. Whether SSL representations internally encode the same acoustic properties as these handcrafted features explicitly measure remains unexplored. Our CCA analysis addresses this question by directly measuring the alignment between WavLM layer representations and these spoofing-relevant acoustic features.
These findings motivate our choice of CCA target features.
\vspace{-0.15cm}
\section{Methodology}
An illustration of our analysis pipeline is shown in Figure~\ref{fig:methodology}.
It comprises two components: 
(i)~per-layer linear probing for discriminability assessment (bottom right, pink block); 
(ii)~canonical correlation analysis (CCA) which is applied between layer representations and handcrafted acoustic features so as to ground representations in interpretable spoofing-relevant utterance characteristics (bottom left, blue block).  
We describe both in the following after a description of the dataset and attack taxonomy and our use of WavLM representations.
\vspace{-0.2cm}
\subsection{Dataset and attack taxonomy}
All work reported in this paper was performed using the ASVspoof 5 database~\cite{wang2025asvspoof5db}, using the Track~1 protocol for spoofing/deepfake detection and with the standard train, development (dev) and evaluation (eval) partition splits~\cite{evalPlan}. 
The train set contains attacks A01--A08 (GlowTTS, GradTTS, FastPitch, VITS variants), the dev set A09--A16 (ToucanTTS, Tacotron2, StarGANv2-VC, YourTTS, VAE-GAN, and others), and the eval set A17--A32 (including ZMM-TTS, MaryTTS, DiffVC, among others, and adversarial methods such as Malafide and Malacopula). 
For analysis, we group the set of 32 attacks into three families, specifically those generated using TTS and VC and those augmented\footnote{For the ASVspoof~5 database, adversarial attacks are applied to already-spoofed utterances generated using either TTS or VC.  There is no application of adversarial attacks to bonafide utterances.  See~\cite{wang2025asvspoof5db} for details.} by adversarial attacks (ADV).

\subsection{WavLM representations and linear probing}

We use the WavLM Base model~\cite{chen2022wavlm} as a frozen feature extractor extracting representations from all layers (L0--L12). 
Unlike wav2vec2.0~\cite{baevski2020wav2vec}, which is trained solely with a contrastive objective using clean speech, WavLM incorporates an additional denoising objective that exposes the model to noisy and overlapped speech during pre-training. 
This design makes WavLM particularly relevant to our analysis, since spoof detection depends on both acoustic distortions and speaker-related properties.
We use the Base variant rather than the Large model for computational efficiency and because its 13-layer architecture provides a tractable setting for layer-wise analysis while still achieving competitive performance for many speech-related tasks.

Since all WavLM layers produce 768-dimensional frame representations, direct layer-wise comparison is straightforward. 
For each layer, we compute mean and standard deviation pooling over the time frame axis, yielding two 768-dimensional statistics which are concatenated into a single 1536-dimensional utterance-level vector suitable for classification.
Using training data only, we learn distinct linear classifiers using the representations extracted from each layer to distinguish bona fide from spoofed utterances. 
We select the best model according to that which produces the lowest equal error rate (EER) for the development set and report EERs for all bona fide and spoofed data in both the development and evaluation sets. 
This approach provides a transparent measure of layer-wise discriminability without the use of complex back-ends obscuring interpretation.

\subsection{CCA-based acoustic feature analysis}                              To ground WavLM representations in interpretable acoustic properties, we estimate the correlation between the representations of each layer and a set of 17 handcrafted acoustic features organised into four categories:   
\vspace{-0.1cm}
\begin{itemize}
\item \textbf{Voice quality (3):} jitter, shimmer and the harmonic-to-noise ratio (HNR), measures of vocal fold micro-perturbations known to be distorted by TTS vocoders~\cite{gupta2025pitchimperfect}.
\item \textbf{Prosodic (4):} F0 mean, F0 standard deviation, energy mean, onset strength.
\item \textbf{Spectral (6):} MFCC mean, spectral centroid, spectral roll-off, ZCR, spectral entropy, spectral flux, general signal descriptors used in CM~\cite{sahidullah2015comparison}.
\item \textbf{Harmonic/formant (4):} harmonic ratio, F1 mean, F2 mean, F3 mean, formant trajectories expose voice conversion artifacts~\cite{cuccovillo2024formant,borrelli2025forensic}.
\end{itemize}
\vspace{-0.1cm}
We use canonical correlation analysis (CCA) to measure how well each WavLM layer's representations align with a set of handcrafted acoustic features~\cite{morcos2018pwcca}. CCA identifies linear projections of two variable sets that are maximally correlated. Here, these correspond to the pooled layer representation $\mathbf{x} \in \mathbb{R}^{d_1}$ and the acoustic feature vector $\mathbf{y} \in \mathbb{R}^{d_2}$, yielding coefficients $\rho_1 \geq \rho_2 \geq \cdots \geq \rho_k$, where $k = \min(d_1, d_2)$.

Standard CCA can be sensitive to directions with low variance. 
We therefore adopt projection-weighted CCA (PWCCA)~\cite{morcos2018pwcca}, which computes a weighted mean of canonical correlations:
  \begin{equation}
  \text{PWCCA} = \frac{\sum_{i=1}^{k} \alpha_i \rho_i}{\sum_{i=1}^{k} \alpha_i}
  \end{equation}
where $\alpha_i$ is the sum of absolute projection weights onto the $i$-th canonical direction, giving greater weight to directions that capture greater variance in the original representation. 
CCA requires the number of samples to exceed the dimensionality of both inputs; following~\cite{morcos2018pwcca}, we reduce layer representations to 50 dimensions via principal component analysis (PCA) prior to computing PWCCA.

We compute correlations between each 1536-dimensional layer representation and the 17 acoustic features matrix.
Canonical correlations are weighted by projection magnitudes and symmetrised by averaging in both directions. 
To report per-feature correlations (Fig.~\ref{fig:cca_heatmap}), we compute the squared correlation between each individual feature and the first canonical variate of the layer representation. Analyses use the train+dev splits only, with the eval set reserved for final reporting. For computational efficiency, we randomly sample up to 5000 speaker-balanced utterances per attack group. This sample size comfortably exceeds the reduced dimensionality and supports reliable correlation estimates.

\vspace{-0.15cm}
\section{Results}

We report results for both development and evaluation sets. 
The study of results for both sets allows us to examine whether layer-wise patterns observed for more familiar attacks generalise to unseen conditions, and to identify where generalisation fails.

\begin{figure*}[t]
\centering
\includegraphics[width=0.88\textwidth]{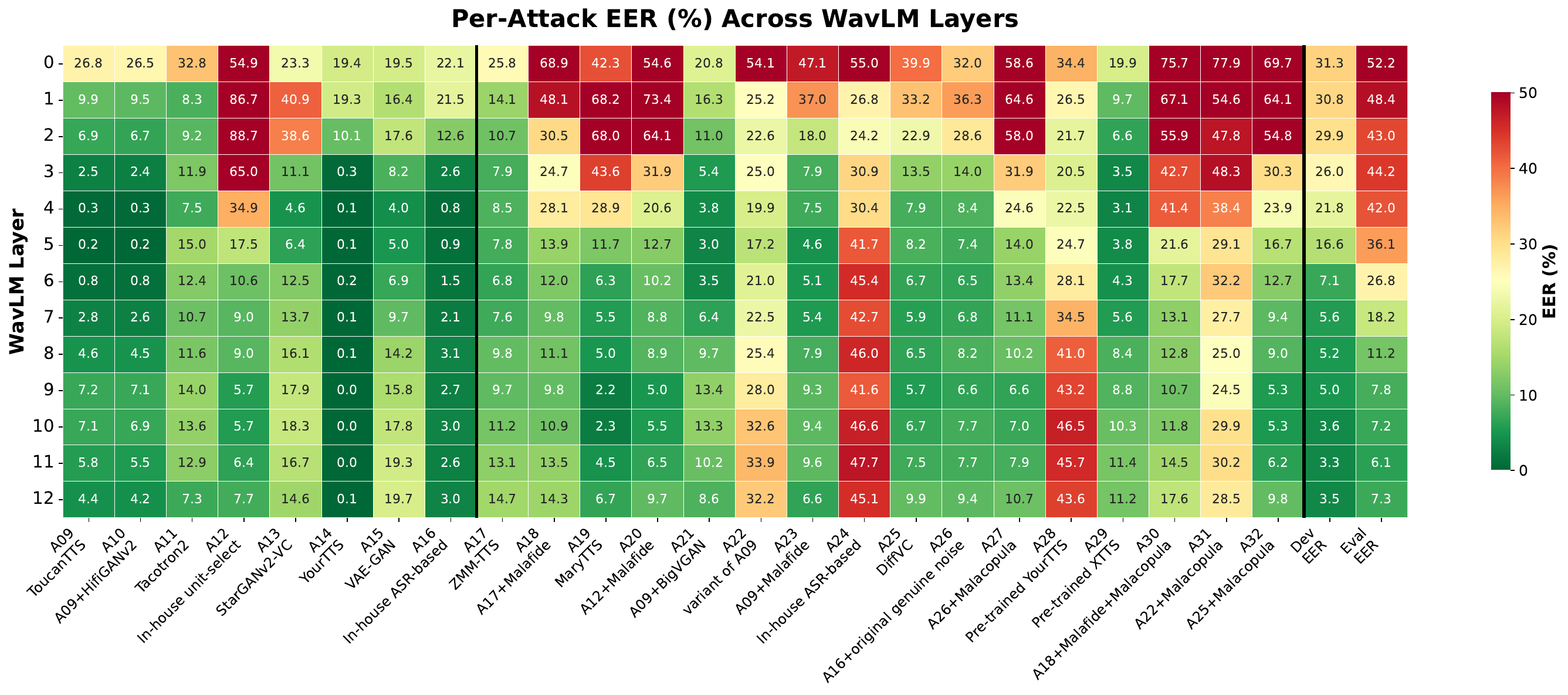}
\vspace{-0.2cm}
\caption{Per-attack EER (\%) across WavLM layers for the dev (A09--A16) and eval (A17--A32) splits, with overall per-layer EER for each set shown in the two rightmost columns. Green indicates low EER (easy detection), whereas red indicates high EER (hard detection). The vertical black line separates dev and eval attacks.}
\label{fig:attack_heatmap}
\end{figure*}

\vspace{-0.2cm}
\subsection{Per-layer probing}

Per-layer EERs from frozen WavLM representations with linear classifiers are shown in Fig.~\ref{fig:attack_heatmap} for both dev and eval sets. 
For both sets, performance generally improves with depth up to L11 resulting in the lowest dev EER of 3.33\% and the lowest eval EER of 6.07\%.
EERs for L12 are higher, showing a loss of discriminative information.
 
A visualisation of EER results for each layer and each attack is shown in Figure~\ref{fig:attack_heatmap}.
Several patterns emerge.
For the dev set, near-zero EERs (dark green colours) are achieved for some TTS attacks (A09, A10, A14 and A16) from mid layers onward. 
Notably, the EER for A12 (unit selection) is higher for early layers but drops sharply after L6 (from up to $\approx$80\% for early layers dropping to $\approx$5\% for late layers), suggesting that deeper representations capture artifacts that shallow layers miss.
For all other attacks in the dev set, and with only few exceptions, EERs are higher than 5\% for all layers.
For the eval set, most attacks follow a similar pattern of decreasing EER with depth. However, for three TTS attacks, A22 (ToucanTTS variant), A24 (ASR-based TTS), and A28 (pre-trained YourTTS), we observe the opposite trend: EERs exceed $\sim$30\% in later layers, higher than that for mid-layers. What distinguishes these attacks from others that become easier to detect with depth is explored further in Section~\ref{sec:cca}.

For attacks augmented with Malafide~\cite{panariello2023malafide} (A18, A20, A23) and Malacopula~\cite{todisco2024malacopula} (A27, A30--32) adversarial filtering, EERs are consistently high (EER$>$20\%) for all layers, with either flat or inverted layer trends.
This is expected since adversarial filters are optimised to suppress detectable artifacts, whether by maximising CM scores (Malafide) or matching target speaker embeddings (Malacopula). 
The uniform failure across all layers suggests that adversarial optimisation successfully removes detectable artifacts regardless of representation depth, exposing a fundamental limitation of artifact-based detection.
\vspace{-0.1cm}
\subsection{CCA analysis}\label{sec:cca}
\begin{figure*}[t]
\centering
\includegraphics[width=0.75\textwidth]{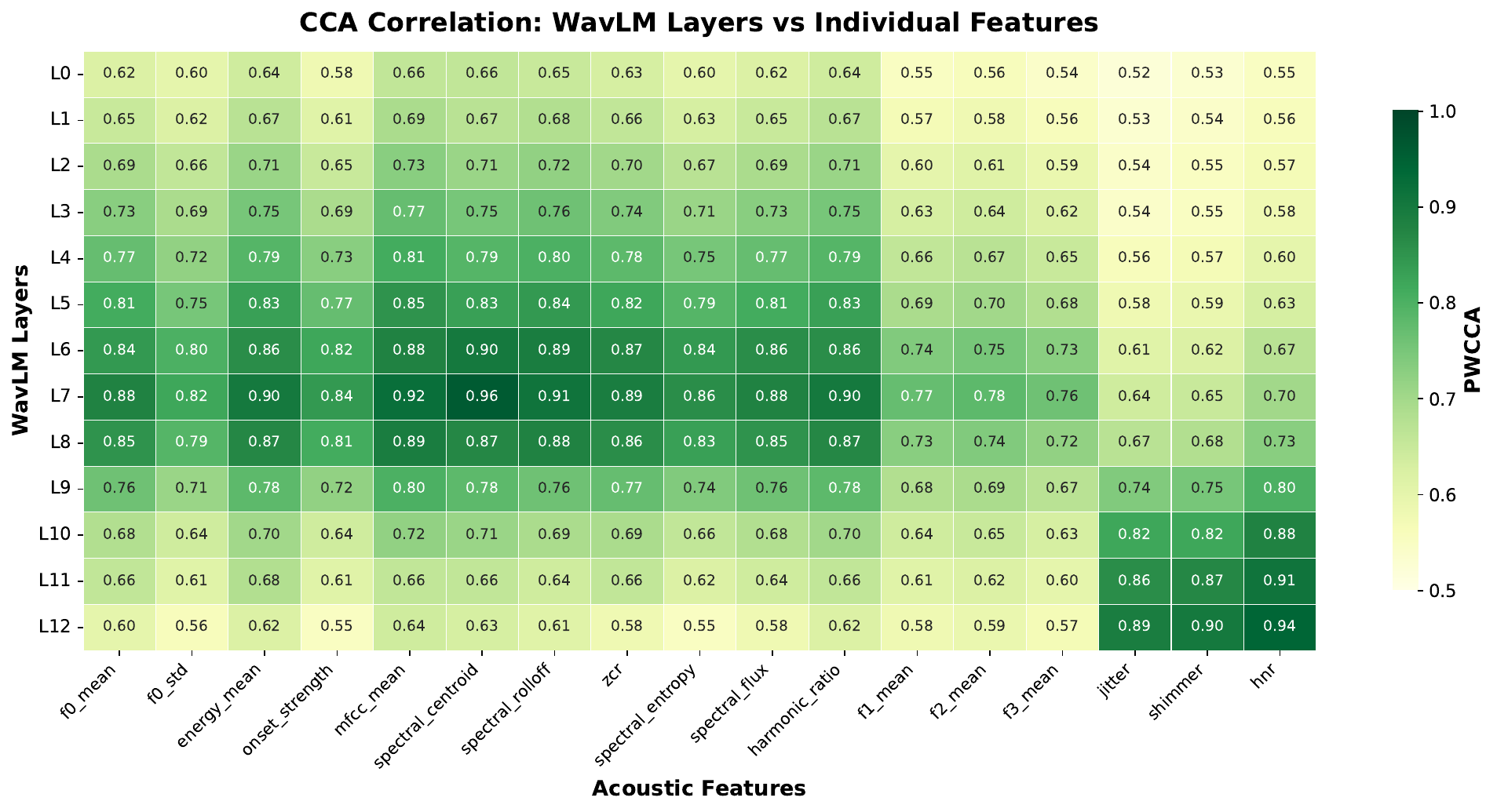}
\vspace{-0.5cm}
\caption{PWCCA correlation between WavLM layers and individual acoustic features. Most features peak at mid-layers and decay, while voice quality measures (jitter, shimmer, HNR) increase with depth.}
\label{fig:cca_heatmap}
\end{figure*}

\vspace{-0.15cm}
PWCCA results are illustrated by a heatmap shown in Figure~\ref{fig:cca_heatmap}.
Colours indicate the correlation between representations extracted from each WavLM layer and individual acoustic features. Two distinct patterns emerge.\\
\textbf{Correlation Peak Around L7-} Most acoustic features follow a consistent pattern: correlation increases from L0, peaks at around L7, and then decays toward L12. Spectral features show the strongest mid-layer alignment, with that for the spectral centroid reaching 0.96 at L7, followed by MFCC mean (0.92) and spectral roll-off (0.91). Prosodic features (F0, energy) follow the same trend with slightly lower peak values. Formant features (F1--F3) show the weakest correlations overall (peaks of 0.70--0.78), suggesting that formant information is less accessible from WavLM representations than spectral or prosodic properties. This pattern mirrors the acoustic-to-abstract transitions reported previously for SSL models~\cite{pasad2021layerwise}: mid-layers encode signal-level properties while later layers encode more abstract representations.\\
\textbf{Uniform Increase in Correlation-} In contrast, jitter, shimmer, and HNR show monotonically increasing correlation with layer depth. Jitter rises from 0.52 (L0) to 0.89 (L12), shimmer from 0.53 to 0.90, and HNR from 0.55 to 0.94. These features measure micro-perturbations in pitch period, amplitude, and harmonic structure properties that vocoders and TTS systems are known to distort ~\cite{gupta2025pitchimperfect}.

These two patterns help explain the probing results. The EER improvements from early to mid-layers coincide with stronger spectral and prosodic encoding, while the further gains at late layers are driven by increasing voice quality sensitivity — even as spectral alignment decays.
Voice quality correlation is greatest for L12, yet EERs are slightly worse than for L11 (7.3\% vs.\ 6.07\%), suggesting a greater balance between retained spectral information and voice quality sensitivity.

\vspace{-0.2cm}
\subsection{Layer subset search}
The per-layer results show clear variation in discriminability across layers, raising the question of whether a compact subset can match or outperform full-layer pooling. To test this, we perform an exhaustive grid search over all $\sum_{k=1}^{13}\binom{13}{k} = 8191$ possible layer combinations. For each subset of $k$ layers, we concatenate the corresponding 1536-dimensional representations into a single $k \times 1536$-dimensional vector and train a linear probe. Subsets are selected on dev only; eval is used once for final reporting.\\ Table~\ref{tab:gridsearch} compares the top subsets against the full 13-layer baseline (19\,968-dimensional input). The best 5-layer subset (L7--9--10--11--12) achieves 6.1\% eval EER, outperforming the full baseline (8.08\%). All top subsets concentrate in the L6--L12 range. This is consistent with the CCA findings in Section~\ref{sec:cca}: layers beyond L6 exhibit the strongest alignment with spoofing-relevant acoustic properties, particularly the voice quality features (jitter, shimmer, HNR) whose correlation increases monotonically with depth. Early layers, which encode primarily low-level spectral information with weaker discriminative relevance, do not contribute complementary information in the linear pooling setting.

\begin{table}[t]
\centering
\caption{Top layer subsets from grid search vs.\ all-layer pooling.}
\label{tab:gridsearch}
\begin{tabular}{lccc}
\toprule
Layers & $k$ & Dev EER & Eval EER \\
\midrule
7--9--10--11--12 & 5 & 2.06\% & \textbf{6.1\%} \\
6--8--9--10--12  & 5 & 2.16\% & 7.16\% \\
6--9--10--11--12 & 5 & 2.02\% & 7.25\% \\
\midrule
All (0--12)      & 13 & 4.59\% & 8.08\% \\
\bottomrule
\end{tabular}
\end{table}
\vspace{-0.3cm}
\subsection{t-SNE Visualisation}
\begin{figure*}[t]
\centering
\includegraphics[width=0.84\textwidth]{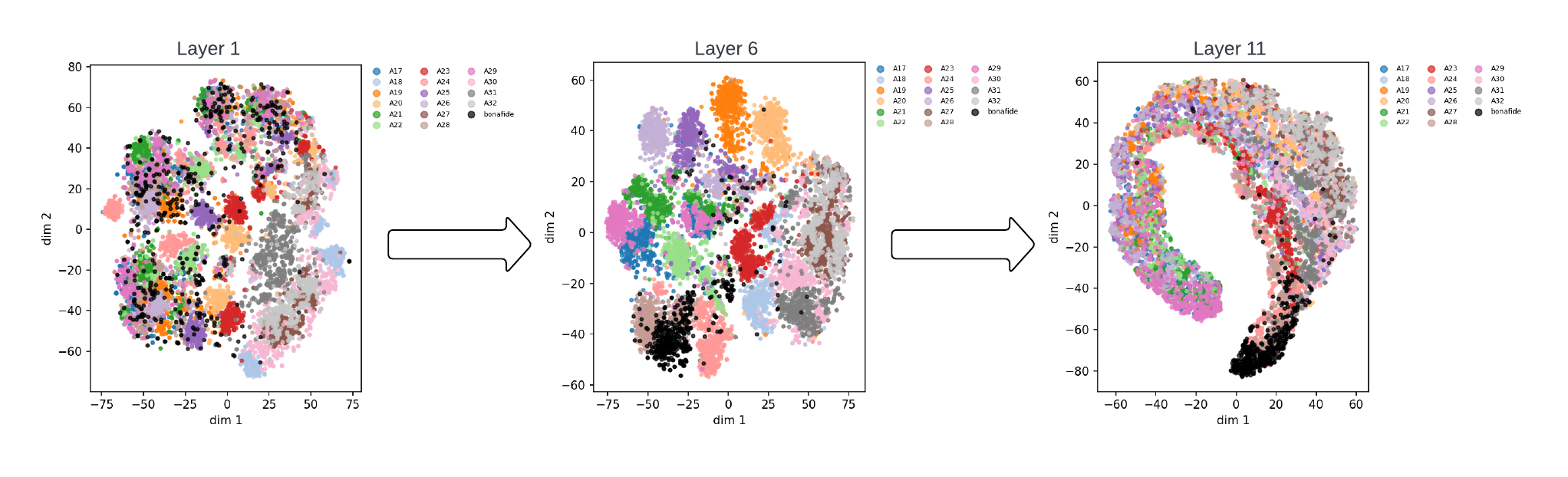}
\vspace{-0.6cm}
\caption{t-SNE visualisation of WavLM representations at layers 1, 6, and 11. Representations transition from overlapping clusters (L1) to discrete attack-specific groupings (L6) to a continuous horseshoe manifold (L11).}
\label{fig:tsne}
\end{figure*}

Figure~\ref{fig:tsne} shows t-SNE projections of WavLM representations at three depths for the eval splits. 
For L1, attacks form overlapping clusters with no clear separation. 
For L6, discrete attack-specific clusters appear and the model appears to distinguish attacks (coloured points) from bona fide utterances (black points).
For L11, attacks form a continuous horseshoe arc with bona fide samples lying at one end of the shoe.
We observed this pattern to be consistent across multiple t-SNE random seeds and perplexity settings (20, 30, 50).

The horseshoe arc is often associated with data organized along a dominant continuous latent dimension~\cite{diaconis2008horseshoes}. 
This observation aligns with the PWCCA result: mid-layers show stronger correlation with categorical spectral features (forming discrete clusters), while late layers correlate more strongly with voice quality features (jitter, shimmer, HNR). 
The shift from discrete to continuous organisation suggests that late layers organise attacks along a gradient of vocal naturalness rather than by categorical attack identity.

\section{Discussion}
\vspace{-0.2cm}
The results offer insights into how WavLM encodes information relevant to spoofing detection.

\noindent\textbf{Voice quality and layer depth -}
The most striking finding from the CCA analysis is that voice quality features (jitter, shimmer, HNR) show increasing correlation with late layers, in contrast to the decay observed for spectral and prosodic features. 
This difference may reflect the WavLM denoising objective during pre-training which exposes the model to degraded and overlapped speech. 
Being tasked with predicting masked tokens from corrupt input, the model may develop sensitivity to fine-grained signals. 
These layers achieve the lowest EER suggesting a direct link: the ability to encode voice quality patterns translates to spoofing discriminability, since vocoders are known to distort these properties~\cite{gupta2025pitchimperfect}.

\noindent\textbf{Attack-specific layer signatures -}
The per-attack heatmap (Fig.~\ref{fig:attack_heatmap}) reveals that detection performance is not uniformly distributed across layers. 
Some attacks (A09, A10, A14, A16) are detected from mid-layers onward, while detection of A12 (unit selection) is better for late layers. 
On the other hand A22, A24 and A28 evade detection at late layers despite being detectable at mid-layers. 
This implies that different synthesis methods leave artifacts at different levels of abstraction: conventional TTS systems may introduce voice quality distortions caught by late layers, while high-quality neural TTS systems trained on greater quantities of data  may produce natural voice quality but leave spectral or prosodic artifacts that are captured by mid layer representations. 
These attack-specific layer signatures have practical implications: a fixed single-layer will at some point fail on certain attack families, motivating multi-layer or adaptive layer selection strategies.

\noindent\textbf{Adversarial robustness -}
The uniform failure across all layers for adversarial attacks (Malafide, Malacopula) highlights a fundamental limitation of artifact-based detection. These methods are explicitly optimised to remove detectable cues, and our results confirm they succeed regardless of layer depth. This suggests that robustness to adversarial attacks may require defenses beyond representation learning, such as adversarial training or detection methods that do not rely solely on acoustic artifacts.

\noindent\textbf{Limitations -}
Linear probes capture only linearly accessible information. We study a single SSL model (WavLM Base), and other architectures may show different layer-wise behaviour. Our CCA analysis also assumes linear relationships between representations and acoustic features. In addition, all experiments are conducted on ASVspoof~5, so the reported trends may not transfer unchanged to other corpora, real-world recordings, or unseen attack families. Finally, our acoustic feature set is not exhaustive.

\section{Conclusions}
We present a layer-wise interpretability study of WavLM for spoof detection using the ASVspoof~5 database combining linear probing, CCA-based acoustic grounding and layer subset search.
Mid-to-late layers (L9--L11) are most discriminative, and a compact 5-layer subset outperforms full 13-layer pooling. 
CCA reveals that voice quality features (jitter, shimmer, HNR) correlate increasingly with late layers, the same layers that achieve the most reliable detection, while spectral and prosodic features peak at mid-layers and decay. 
t-SNE visualisations corroborate these observations, showing that late layers organise attacks along a continuous gradient of vocal naturalness. 
These findings provide an interpretable grounding for SSL-based layer selection and highlight the challenge posed by adversarial attacks that are designed to evade detection.
Future work should examine whether these layer-wise trends remain stable across other corpora, real-world conditions, and unseen attack families.

\balance
\bibliographystyle{IEEEtran}
\bibliography{references}

\end{document}